\documentclass{Interspeech2024}

\interspeechcameraready 
\usepackage{amsmath}
\usepackage{amssymb}
\usepackage{mathtools}
\usepackage{amsthm}
\usepackage{hyperref}
\usepackage{url}            
\usepackage{booktabs}       
\usepackage{amsfonts}       
\usepackage{nicefrac}       
\usepackage{microtype}      
\usepackage{xcolor}         
\usepackage{adjustbox}
\usepackage{multirow}
\usepackage{bbm}
\usepackage[euler]{textgreek}
\usepackage{pifont}
\usepackage{svg}
\usepackage{wrapfig}

\title{AVCap: Leveraging Audio-Visual Features as Text Tokens for Captioning}

\name[affiliation={1}]{Jongsuk}{Kim}
\name[affiliation={1}]{Jiwon}{Shin}
\name[affiliation={1}]{Junmo}{Kim}

\address{
  $^1$Korea Advanced Institute of Science and Technology, South Korea}
\email{\{jskpop, tlswldnjs030, junmo.kim\}@kaist.ac.kr}


\begin{document}

\maketitle

\begin{abstract}
In recent years, advancements in representation learning and language models have propelled Automated Captioning (AC) to new heights, enabling the generation of human-level descriptions. 
Leveraging these advancements, we propose \textbf{AVCap}, an \textbf{A}udio-\textbf{V}isual \textbf{Cap}tioning framework, a simple yet powerful baseline approach applicable to audio-visual captioning.
AVCap utilizes audio-visual features as text tokens, which has many advantages not only in performance but also in the extensibility and scalability of the model.
AVCap is designed around three pivotal dimensions: the exploration of optimal audio-visual encoder architectures, the adaptation of pre-trained models according to the characteristics of generated text, and the investigation into the efficacy of modality fusion in captioning.
Our method outperforms existing audio-visual captioning methods across all metrics and the code is available on \url{https://github.com/JongSuk1/AVCap}.
\end{abstract}

\section{Introduction}
Recent advancements in representation learning and language models have led to significant improvements in the performance of Automated Captioning (AC). These advances have enabled the generation of descriptions for auditory, visual, and other modalities without domain constraints.

One of the key components in a captioning framework is how information from the input data is conveyed to the text decoder. Considering a transformer-based text decoder, there are two main streams for utilizing input data: cross-attention and self-attention. 
The cross-attention-based method, where input features serve as keys and values, shows remarkable performance and led to significant research interest~\cite{liu23l_interspeech, xiao2022local, mei2021audio, gontier2021automated}. 
On the other hand, the self-attention-based method takes input data features to text tokens, incorporating them into the input of the text decoder. 
This approach enhances the model's scalability, adapts seamlessly to increased data volumes, and offers flexibility for expansion into various tasks.
Consequently, it has proven high effectiveness not only in captioning but also in applications such as visual question answering~\cite{wang2022git} and visual instruction tuning~\cite{liu2024visual}, showcasing its versatility.

Meanwhile, leveraging the weights of models pre-trained with large-scale unlabeled datasets in a self-supervised manner has emerged as a highly effective strategy across various downstream tasks.
Recent advances in self-supervised multi-modal representation learning, utilizing pretext tasks such as masked modeling~\cite{gong2022contrastive, georgescu2023audiovisual} or contrastive learning~\cite{gong2022contrastive, guzhov2022audioclip}, has demonstrated its superior ability for encoding multi-modal inputs into embeddings that contain rich semantics.
This approach is particularly pivotal in multi-modal captioning, where the ability to effectively encode and integrate features from different domains is essential.
However, effectively utilizing the pre-trained model in the multi-modal captioning task is relatively unexplored.

\begin{figure}
    \centering
    \includegraphics[width=\linewidth]{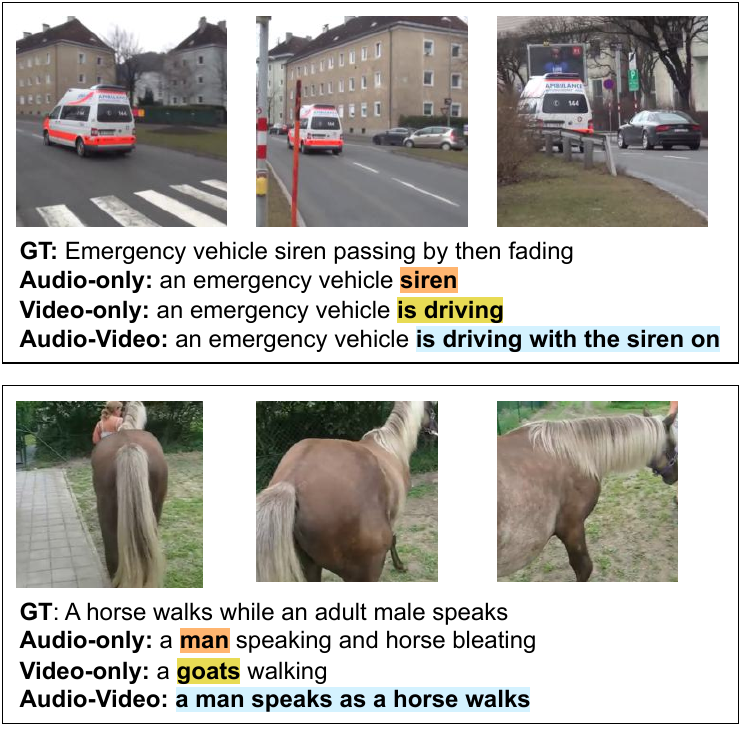}
    \vspace{-1.5em}
    \caption{Qualitative results comparing captions generated from audio-only, video-only, and audio-video training.}
    \label{fig:intro}
    \vspace{-2em}
\end{figure}

Along this line, we present an \textbf{AVCap}, \textbf{A}udio-\textbf{V}isual \textbf{Cap}tioning method that leverages audio-visual features as text tokens with self-attention. Our analysis is structured around three critical dimensions.
First, we explore the optimal audio-visual encoder architecture to effectively perform multi-modal captioning. This involves comparing the performance of various base architectures commonly used in multi-modal representation learning.
Second, we discover the properties of the generated texts vary depending on the adaptation method of the pre-trained model. Thus, we design an appropriate adaptation strategy according to the characteristics of the text to be generated.
Lastly, we investigate the effect of modality fusion in captioning. 
As shown in Figure~\ref{fig:intro}, single-domain captioning has limitations in that it cannot represent all of the given circumstances.
On the other hand, the integration of both audio and visual information enables more precise description.

To the best of our knowledge, our method is the first approach utilizing audio-visual features as text tokens in audio-visual captioning tasks. AVCap outperforms previous audio-visual captioning methods in various evaluation metrics, and the extensive ablation studies verify the efficacy of our approach.

\begin{figure*}
    \centering
    \includegraphics[width=\textwidth]{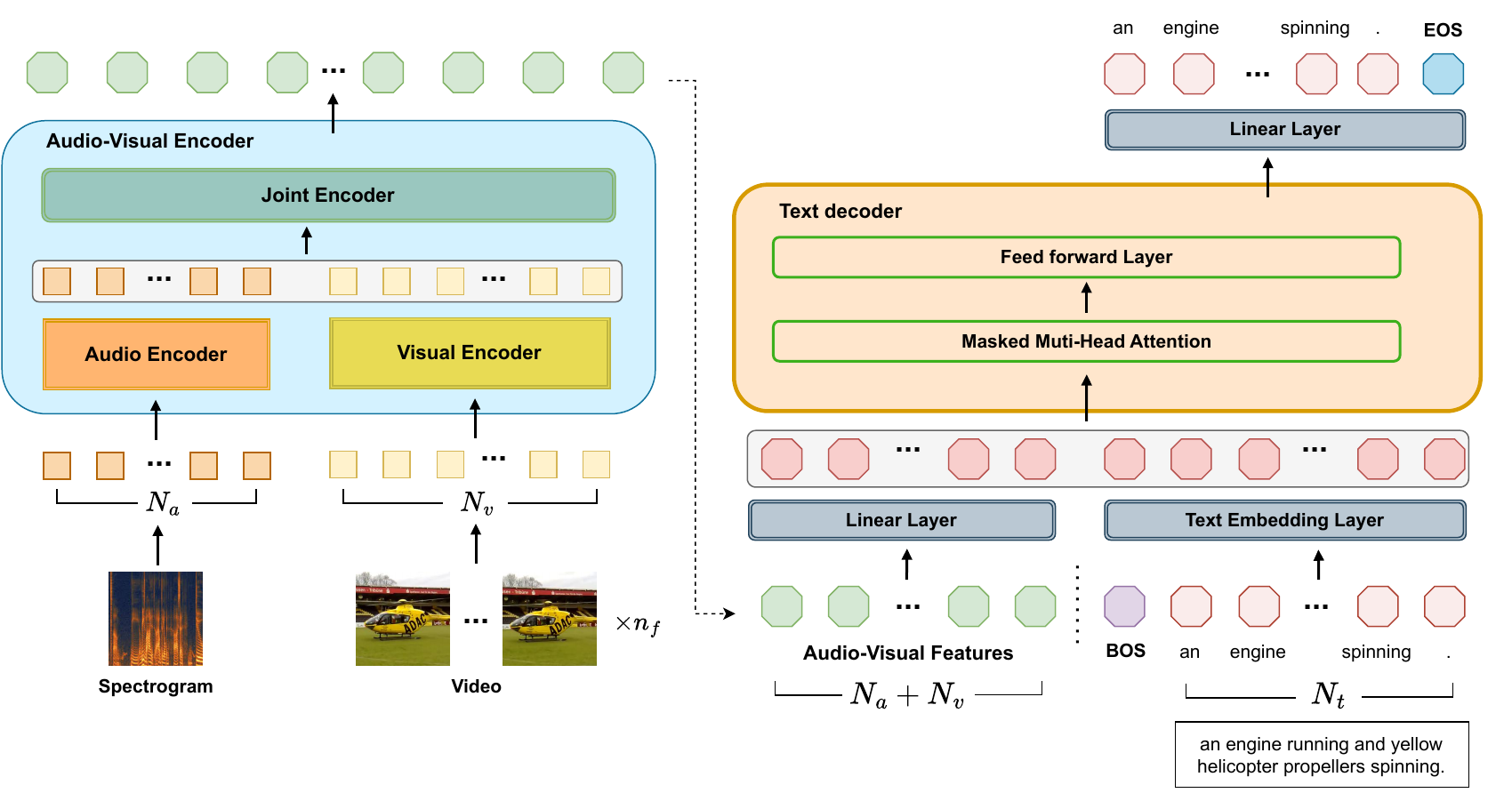}
    \caption{Overview of AVCap.}
    \label{fig:overview}
    \vspace{-1em}
\end{figure*}

\section{Related Works}
\subsection{Audio Captioning}
The audio captioning generally adopted an encoder-decoder framework~\cite{drossos2017automated,wu2019audio,eren2020audio}, where the encoder derives latent representations of the audio features from the input, and the decoder utilizes the representations subsequently. 
To enhance the performance of the audio captioning model, not only varying network architectures~\cite{xiao2022local,mei2021audio,chen2022interactive,koh2022automated} but also various approaches have been introduced, such as applying contrastive learning with contrastive loss~\cite{liu2021cl4ac}, reinforcement ~\cite{xu2020crnn,mei2021encoder}, using auxiliary information~\cite{sun2022automated,koizumi2020transformer}, and other modalities~\cite{xu2017learning}.
\vspace{-0.5em}
\subsection{Audio-Visual Captioning}
Additionally, research has been conducted on captioning by encoding audio-visual data together. Visually-aware audio captioning enables precise differentiation of ambiguous sounds. 
VACT~\cite{liu23l_interspeech} employed a cross-attention technique, which utilizes both audio and video features as key, values in the attention layer. 
MDVC~\cite{iashin2020multi} proposed a dense video captioning method by leveraging audio and visual information.
Another recent study~\cite{chen2022interactive} revealed the role of audio in video captioning using pre-trained model leveraging video and audio modalities. 

\subsection{Audio-Visual Representation Learning}
Audio-visual representation learning plays a crucial role in captioning by understanding the interactions between modalities. 
Contrastive learning and masked modeling stand as typical instances of representation learning.
AudioCLIP~\cite{guzhov2022audioclip} extends CLIP to align the feature space of audio-visual-text data, showcasing powerful performance across various downstream tasks.
AudioVisualMAE~\cite{georgescu2023audiovisual} explored a straightforward expansion of Masked Autoencoders (MAE) designed for audio-visual representation learning. 
Furthermore, CAV-MAE~\cite{gong2022contrastive} and MAViL~\cite{huang2024mavil} merged contrastive learning with masked data modeling techniques to train audio-visual representations.

\section{Method}
Audio-visual captioning aims to generate text caption by taking audio data $a$ and visual data $v$ as input.
Our framework comprises three main parts: (1) encoding audio-visual data, (2) projecting and concatenating audio-visual embeddings with text embeddings, and (3) decoding text. Each component of our method is explained in the following sections.
\subsection{Pre-Processing}
To train the model, we use the Audiocaps~\cite{kim2019audiocaps} dataset.
This dataset contains 10 seconds of video with sound, and corresponding captions describing each video. 
The audio, video, and text are pre-processed for training, as described in detail below.
\newline
\textbf{Audio.} 
The audio waveform is transformed into a series of 128-dimensional log Mel filterbank features. These features are derived using a 25ms Hanning window applied at 10ms intervals, producing a spectrogram with dimensions of $1024 \times 128$. The spectrogram is split into $16\times16$ patches, resulting in a sequence $x_a=[a_1,...,a_{N_a}]$, where $N_a=512$.
\newline
\textbf{Video.} 
The video is sampled at 2 frames per second to obtain 20 RGB images. 
Among them, a subset of $n_f$ consecutive frames is selected to construct the video input with dimension $3\times n_f \times 224 \times 224$.
The video input is split into $2\times 16\times 16$ blocks to form the sequence $x_v=[v_1,...,v_{N_v}]$, where $N_v = 196\times n_f/2$.
In cases where only a single frame is used (i.e. $n_f=1$), we apply the two-dimensional split to construct a visual token sequence and $N_v=196$.
\newline
\textbf{Text.} 
The tokenizer converts the captions into $N_t$ text tokens. During the training phase, these tokens are utilized to construct the input and target. 
The text input $x_t$ is prefixed with the \texttt{[BOS]} token to indicate the start of the sentence, and the target $y_t$ is suffixed with the \texttt{[EOS]} token to indicate the end of the sentence as $x_t=[\texttt{[BOS]},t_1,...,t_{N_t}]$ and $y_t=[t_1,...,t_{N_t},\texttt{[EOS]}]$. 
Then, the text input $x_t$ passes through an embedding layer to obtain the text embedding $h_t\in \mathbb{R}^{(N_t+1)\times D}$, where $D$ denotes the embedding dimension.
Note that $N_t$ varies depending on the caption, thus in a multi-batch environment, each tokenized text is padded to the length of the longest text token sequence in the batch.

\subsection{Encoding Audio-Visual Embeddings}
To encode audio-visual input data, we combine the modality-specific encoders and joint encoder, which are widely used in audio-visual representation learning. As shown in Figure~\ref{fig:overview}, input pair $(x_a,x_v)$ are independently encoded with modality-specific information, and then passed through a joint encoder to produce a feature that fuses the information from both domains.
Based on the Vision Transformer (ViT)~\cite{dosovitskiy2020image}, the encoder is configured using Multi-head Self Attention (MSA) layers, MLP blocks, and Layer Normalization (LN), and the process of encoding audio information can be shown as follows.
\begin{align}
    & h_0 = [a_1\mathbf{E}_a;a_2\mathbf{E}_a;...;a_{N_a}\mathbf{E}_a] + \mathbf{E}_{a,pos} \nonumber\\
    & h'_l = \operatorname{MSA}(\operatorname{LN}(h_{l-1})) + h_{l-1} &l = 1,...,L \nonumber\\
    & h_l = \operatorname{MLP}(\operatorname{LN}(h'_l)) + h'_l &l=1,...,L  \nonumber\\
    & h_a =\operatorname{LN}(\operatorname{MeanPool}(h_L)),
\end{align}
where $\mathbf{E}_a\in \mathbb{R}^{{p^2}\times D}$ is linear projection layer with patch size $p=16$ and $\mathbf{E}_{a,pos}\in \mathbb{R}^{N_a\times D}$ denotes positional embedding. The video data is encoded using the same method to obtain feature $h_v$ with an identical number of layers. 

Afterward, the audio and visual features are concatenated and passed through a joint encoder to obtain $z_{av}$, utilizing $S$ stacked layers in the joint encoder as below.
\begin{align}
    & z_{0} = \operatorname{Concat}(h_a, h_v) \nonumber\\
    & z'_{s} = \operatorname{MSA}(\operatorname{LN}(z_{s-1})) + z_{s-1} & s = 1,...,S \nonumber\\
    & z_{s} = \operatorname{MLP}(\operatorname{LN}(z'_s)) + z'_s & s=1,...,S \nonumber\\
    & z_{av} =\operatorname{LN}(z_S).
\end{align}

\subsection{Decoding Audio-Visual-Text Embeddings}
In our framework, the text decoder takes text and audio-visual embedding as input and generates text output.
We first project the audio-visual feature $z_{av}$ into the text embedding space, as follows:
\begin{equation}
    h_{av\rightarrow t} = W_t\cdot z_{av} + b_t.
\end{equation}
Thus, audio-visual tokens in the text embedding space can be obtained. 
These tokens are concatenated with the text embedding $h_t$ and pass through the text decoder. 
\setlength{\intextsep}{0pt}
\setlength{\columnsep}{10pt}
\begin{wrapfigure}{r}{0.5\linewidth}
    \centering
    \includegraphics[width=\linewidth]{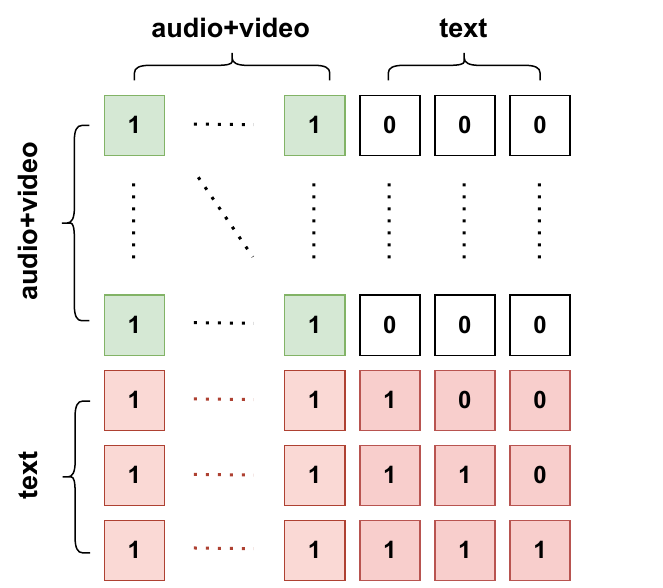}
    \vspace{-1.5em}
    \caption{Attention mask $M$.}
    \label{fig:mask}
\end{wrapfigure}
The text decoder operates with the attention mask $M$, which enables audio-visual tokens to reference each other, while text tokens refer only to preceding tokens and audio-visual tokens, as illustrated in Figure~\ref{fig:mask}. Note that $(i,j)=1$ means the $i$-th output reflects the $j$-th input, while other implies $i$-th output and $j$-th input is independent.
Finally, we can express the process of obtaining the output of the text decoder $g_t$ as below.
\begin{align}
z_t = g_t(\operatorname{Concat}(h_{av\rightarrow t}, h_t); M).
\end{align}

\subsection{Training and Inference}
To train the model, we obtain text description from decoder output $z_t$, given by $\hat{y} = z_t[N_a+N_v:]$.
Specifically, the loss function for the input $a$ and $v$ can be described as follows:
\begin{align}
    \displaystyle \mathcal{L} &= \frac{1}{N_t+1}\sum^{N_t+1}_{i=1} \operatorname{CE}(y_i,\hat{y}_i) \\
    &=\frac{1}{N_t+1}\sum^{N_t+1}_{i=1} \operatorname{CE}\left(y_i, p(y_i|a,v,\displaystyle\bigcup^{i-1}_{j=0}\{y_j\})\right),
\end{align}
where $\operatorname{CE}$ is the cross-entropy loss function.
During inference, we use the beam search~\cite{wu2016google} algorithm with a beam size of 4 and a length penalty of 0.6 to effectively generate captions. 
Note that, for all test audio-visual samples in the inference phase, the text input 
$x_t$ is set to $[\texttt{[BOS]}]$.

\begin{table*}[h]
    \caption{Captioning performance on the AudioCaps test set. \textbf{Bold} represents the best methods within audio-visual-based methods and \underline{underline} means the overall best methods. TD denotes the Text Decoder.}
    \centering
    \begin{adjustbox}{max width=0.95\linewidth}
    \begin{tabular}{lcccccccccc} 
         \hline
         Method & BLEU$_{1}$ & BLEU$_{2}$ & BLEU$_{3}$ & BLEU$_{4}$ & ROUGE$_{L}$ & METEOR & CIDEr & SPICE & SPIDEr \\ 
         \hline
         \footnotesize\textbf{\textit{Audio-Based Models}}\\
         AudioCaps~\cite{kim2019audiocaps} & 0.614 & 0.446 & 0.317 & 0.219 & 0.450 & 0.203 & 0.593 & 0.144 & 0.369 \\
         LHDFF~\cite{sun2023dual} & 0.674 & 0.502 & 0.368 & 0.267 & 0.483 & 0.232 & 0.680 & 0.171 & 0.426 \\
         ACT~\cite{mei2021audio} & 0.685 & 0.518 & 0.376 & 0.263 & 0.488 & 0.233 & 0.678 & 0.169 & 0.424 \\
         BART Tags~\cite{gontier2021automated} & 0.699 & 0.523 & 0.380 & 0.266 & 0.493 & 0.241 & 0.753 & 0.176 & 0.465 \\
         CNN-GPT2~\cite{kim2023prefix} & \underline{0.713} & \underline{0.552} & \underline{0.421} & \underline{0.309} & \underline{0.503} & 0.240 & 0.733 & 0.177 & 0.455 \\
         \hline
         \footnotesize\textbf{\textit{Audio-Visual Based Models}} \\
         V-ACT~\cite{liu23l_interspeech} & 0.698 & 0.527 & 0.388 & 0.281 & 0.494 & 0.237 & 0.711 & 0.172 & 0.442  \\
         AVCap (TD-freeze, Ours) & \textbf{0.708} & \textbf{0.537} & \textbf{0.405} & \textbf{0.295} & \textbf{0.498} & 0.228 &0.744 &0.162 & 0.455 \\
         AVCap (TD-train, Ours) & 0.681 & 0.515 & 0.387 & 0.287 & 0.491 & \underline{\textbf{0.243}} & \underline{\textbf{0.758}} & \underline{\textbf{0.178}} & \underline{\textbf{0.468}} \\
        \hline
    \end{tabular}
    \end{adjustbox}
    \label{tab:results} 
\end{table*}
\section{Experimental Results}
\subsection{Experimental settings}
\subsubsection{Dataset.} 
For training and evaluation, we utilize the AudioCaps~\cite{kim2019audiocaps} dataset. 
The video and audio clip are accessible via a YouTube link and paired with a caption.
Since some links are no longer accessible, we obtained 48,595 out of 49,838 audio-video clips for the train set and 944 out of 975 for the test set. 

\subsubsection{Evaluation Metrics.}
To measure the quality of captions comprehensively, we employ several metrics.
BLEU$_1$ (B@1) to BLEU$_4$ (B@4) assess n-gram precision between generated text and ground truth text. 
ROUGE$_L$ (R) focuses on the longest common subsequence to evaluate recall, and METEOR (M) considers both precision and recall using synonyms and stems.
CIDEr (C) measures consensus among ground truth captions, emphasizing the uniqueness of generated captions. 
SPICE (S) evaluates the semantic accuracy of captions in depicting objects, attributes, and relationships in input data. 
SPIDEr (SC) combines the SPICE and CIDEr to achieve a balance between these two scores.

\subsubsection{Implementation Details.} 
In the training phase, we apply the label smoothing factor of 0.1 to compute loss. 
We use the AdamW optimizer with weight decay at $5\times 10^{-7}$, $\beta_1=0.95$ and $\beta_2=0.999$. 
The learning rate warms up over the first 50 steps to reach the peak learning rate $1\times 10^{-4}$ and then follows the cosine decay until the total of 2500 steps. 
When using pre-trained models, the audio-visual encoder is initialized with a \texttt{CAV-MAE-scale++}\footnote{https://github.com/YuanGongND/cav-mae}, while the text decoder is initialized with a text decoder of \texttt{GIT-base}\footnote{https://github.com/microsoft/GenerativeImage2Text} which follows BERT$_B$~\cite{devlin2018bert} with 12 layers.
\subsection{Main Results}
\label{sec:result}
Table~\ref{tab:results} demonstrates the captioning performance of our model on the AudioCaps dataset. We initialize the audio-visual encoder and text decoder with a pre-trained model while keeping the audio-visual encoder updated during training. 
A notable observation is the incorporation of the pre-trained text decoder leads to distinct trends across two groups of evaluation metrics. 
When freezing the text decoder, the model achieves higher scores in BLEU$_n$ score and ROUGE$_L$, indicating a stronger lexical alignment with ground truth captions. 
In contrast, training the decoder results in performance improvement in the other metrics, reflecting enhanced semantic relevance and informativeness in the generated captions. 
These differences highlight that caption generation may require different strategies depending on the focus between lexical accuracy and semantic content.
\begin{table}[t]
    \centering
    \caption{Architecture Search. $L$ denotes the number of layers in the modality-specific encoder.}
    \label{tab:arch}
    \begin{tabular}{c|cccccc}
        \toprule
        $L$ & B@3   & B@4   & R     & M     & C  & S   \\ 
        \midrule
        0  & 0.289 & 0.189 & 0.412 & 0.182 & 0.539 & 0.127\\
        6  & 0.297 & 0.193 & 0.417 & \textbf{0.187} & 0.537 & \textbf{0.133}\\
        11 & \textbf{0.311} & \textbf{0.217} & \textbf{0.428} & 0.184 & \textbf{0.543} & 0.131 \\
        12 & 0.301 & 0.198 & 0.424 & 0.178 & 0.522 & 0.124\\ 
        \bottomrule
    \end{tabular}
    \vspace{-1em}
\end{table}

Meanwhile, we achieve higher performance than VACT~\cite{liu23l_interspeech}, which leverages audio-visual information through cross-attention. The result indicates that our method serves as a robust baseline approach within the captioning field.
However, our model performs slightly worse than the audio-based model on some metrics, due to the difference in power of the pre-trained language models.
More analysis of the result can be found in Section~\ref{sec:modal}.

\setlength{\intextsep}{0pt}
\setlength{\columnsep}{10pt}
\setlength{\belowcaptionskip}{0pt}
\subsection{Ablation Studies}
In this section, we conduct ablation studies on the various components of our method. For all experiments, the text decoder is consistently initialized with a pre-trained model and frozen.
\subsubsection{Architecture Search.}
Table~\ref{tab:arch} illustrates the performance according to different designs of the audio-visual encoder.
For a fair comparison, we train the model from scratch and keep the number of layers $S$ in the joint encoder to be $L+S=12$.
When $L=0$, the audio and visual patches are concatenated and directly utilized as input to the joint encoder, and $L=12$ indicates the dual encoder structure is used without a joint encoder. 
In most metrics, the best performance is obtained when $L=11$, which is similar to the most common structure in audio-visual representation learning~\cite{gong2022contrastive,georgescu2023audiovisual, huang2024mavil}.
Hence, we use 11 layers in the audio-visual encoder architecture.

\subsubsection{Pre-trained Audio-Visual Encoder Adaptation.}
We conduct experiments on the effective utilization of pre-trained audio-visual encoders in Table~\ref{tab:training-strategies}. 
Unlike the text decoder, the pre-trained audio-visual encoder shows consistently high performance when updating all parameters without freezing.
The results highlight the significant benefits of tuning the encoder through captioning loss to improve overall model performance.
\begin{table}[t]
    \centering
    \caption{Pre-trained audio-visual encoder adaptation strategy.}
    \label{tab:training-strategies}
    \begin{adjustbox}{width=\linewidth}
    \begin{tabular}{l|cccccc}
        \toprule
        AV Encoder      & B@3   & B@4   & R        & M      & C & S\\ \midrule
        train w/ scratch         & 0.311 & 0.217 & 0.428 & 0.184 & 0.543 & 0.131\\
        train w/ PT     & \textbf{0.405} & \textbf{0.295} & \textbf{0.498}    & \textbf{0.228}  & \textbf{0.744}  & \textbf{0.162} \\
        freeze w/ PT    & 0.320 & 0.207 & 0.453    & 0.195  & 0.581 & 0.135\\ 
        \bottomrule
    \end{tabular}
    \end{adjustbox}
\end{table}

\begin{table}[t]
    \centering
    \caption{Effect of Data Modality and $n_f$.}
    \label{tab:modal}
    \begin{adjustbox}{max width=\linewidth}
        \begin{tabular}{ll|cccccc}
            \toprule
            Modality & $n_f$ & B@3   & B@4   & R     & M     & C     & S     \\ 
            \midrule
            V    & 1 & 0.247 & 0.158 & 0.391 & 0.153 & 0.441 & 0.107 \\
            A    & 1 & 0.401 & 0.292 & 0.494 & 0.226 & 0.735 & 0.164 \\
            A+V  & 1 & 0.405 & 0.295 & 0.498 & 0.228 & \textbf{0.744}& 0.162 \\ 
            A+V  & 4  & 0.400 & 0.290 & 0.488 & 0.226 & 0.710 & 0.162 \\
            A+V  & 8  & \textbf{0.407} & \textbf{0.301} & \textbf{0.499} & \textbf{0.230}  & 0.731 & \textbf{0.165} \\ 
            \bottomrule
        \end{tabular}
    \end{adjustbox}
\end{table}
\subsubsection{Effect of Data Modality.}
\label{sec:modal}
Table~\ref{tab:modal} shows the result depending on the input data modalities. 
The model trained using only visual data performs lower than other methods, while the model using only audio data shows comparable results to the model using both audio-visual data. 
We can speculate a reason for the result as the audio-centric nature of the AudioCaps dataset.
This is further supported by the visual information in the input increases with the number of frames, the performance is not significantly affected. 
On the other hand, qualitative results in Figure 1 show that leveraging both audio and visual data provides a more comprehensive description of the paired data. 
\section{Conclusion}
In this work, we propose a straightforward audio-visual captioning method, named AVCap, for levering audio-visual features into text tokens. 
As evidenced by studies aligned with our approach~\cite{liu2024visual}, AVCap demonstrates significant potential to advance the field of audio-visual learning.
We hope that our research can be applied beyond the audio-visual learning field to the realm of general multi-modal learning.
\section{Acknowledgements}
This work was supported by Institute of Information \& communications Technology Planning \& Evaluation (IITP) grant funded by the Korea government(MSIT) (No. 2022-0-00184, Development and Study of AI Technologies to Inexpensively Conform to Evolving Policy on Ethics)
\bibliographystyle{IEEEtran}
\bibliography{mybib}

\begin{thebibliography}{10}
\providecommand{\url}[1]{#1}
\csname url@samestyle\endcsname
\providecommand{\newblock}{\relax}
\providecommand{\bibinfo}[2]{#2}
\providecommand{\BIBentrySTDinterwordspacing}{\spaceskip=0pt\relax}
\providecommand{\BIBentryALTinterwordstretchfactor}{4}
\providecommand{\BIBentryALTinterwordspacing}{\spaceskip=\fontdimen2\font plus
\BIBentryALTinterwordstretchfactor\fontdimen3\font minus \fontdimen4\font\relax}
\providecommand{\BIBforeignlanguage}[2]{{%
\expandafter\ifx\csname l@#1\endcsname\relax
\typeout{** WARNING: IEEEtran.bst: No hyphenation pattern has been}%
\typeout{** loaded for the language `#1'. Using the pattern for}%
\typeout{** the default language instead.}%
\else
\language=\csname l@#1\endcsname
\fi
#2}}
\providecommand{\BIBdecl}{\relax}
\BIBdecl

\bibitem{liu23l_interspeech}
X.~Liu, Q.~Huang, X.~Mei, H.~Liu, Q.~Kong, J.~Sun, S.~Li, T.~Ko, Y.~Zhang, L.~H. Tang, M.~D. Plumbley, V.~Kılıç, and W.~Wang, ``{Visually-Aware Audio Captioning With Adaptive Audio-Visual Attention},'' in \emph{Proc. INTERSPEECH 2023}, 2023, pp. 2838--2842.

\bibitem{xiao2022local}
F.~Xiao, J.~Guan, H.~Lan, Q.~Zhu, and W.~Wang, ``Local information assisted attention-free decoder for audio captioning,'' \emph{IEEE signal processing letters}, vol.~29, pp. 1604--1608, 2022.

\bibitem{mei2021audio}
X.~Mei, X.~Liu, Q.~Huang, M.~D. Plumbley, and W.~Wang, ``Audio captioning transformer,'' \emph{arXiv preprint arXiv:2107.09817}, 2021.

\bibitem{gontier2021automated}
F.~Gontier, R.~Serizel, and C.~Cerisara, ``Automated audio captioning by fine-tuning bart with audioset tags,'' in \emph{DCASE 2021-6th Workshop on Detection and Classification of Acoustic Scenes and Events}, 2021.

\bibitem{wang2022git}
J.~Wang, Z.~Yang, X.~Hu, L.~Li, K.~Lin, Z.~Gan, Z.~Liu, C.~Liu, and L.~Wang, ``Git: A generative image-to-text transformer for vision and language,'' \emph{arXiv preprint arXiv:2205.14100}, 2022.

\bibitem{liu2024visual}
H.~Liu, C.~Li, Q.~Wu, and Y.~J. Lee, ``Visual instruction tuning,'' \emph{Advances in neural information processing systems}, vol.~36, 2024.

\bibitem{gong2022contrastive}
Y.~Gong, A.~Rouditchenko, A.~H. Liu, D.~Harwath, L.~Karlinsky, H.~Kuehne, and J.~Glass, ``Contrastive audio-visual masked autoencoder,'' \emph{arXiv preprint arXiv:2210.07839}, 2022.

\bibitem{georgescu2023audiovisual}
M.-I. Georgescu, E.~Fonseca, R.~T. Ionescu, M.~Lucic, C.~Schmid, and A.~Arnab, ``Audiovisual masked autoencoders,'' in \emph{Proceedings of the IEEE/CVF International Conference on Computer Vision}, 2023, pp. 16\,144--16\,154.

\bibitem{guzhov2022audioclip}
A.~Guzhov, F.~Raue, J.~Hees, and A.~Dengel, ``Audioclip: Extending clip to image, text and audio,'' in \emph{ICASSP 2022-2022 IEEE International Conference on Acoustics, Speech and Signal Processing (ICASSP)}.\hskip 1em plus 0.5em minus 0.4em\relax IEEE, 2022, pp. 976--980.

\bibitem{drossos2017automated}
K.~Drossos, S.~Adavanne, and T.~Virtanen, ``Automated audio captioning with recurrent neural networks,'' in \emph{2017 IEEE Workshop on Applications of Signal Processing to Audio and Acoustics (WASPAA)}.\hskip 1em plus 0.5em minus 0.4em\relax IEEE, 2017, pp. 374--378.

\bibitem{wu2019audio}
M.~Wu, H.~Dinkel, and K.~Yu, ``Audio caption: Listen and tell,'' in \emph{ICASSP 2019-2019 IEEE International Conference on Acoustics, Speech and Signal Processing (ICASSP)}.\hskip 1em plus 0.5em minus 0.4em\relax IEEE, 2019, pp. 830--834.

\bibitem{eren2020audio}
A.~{\"O}. Eren and M.~Sert, ``Audio captioning based on combined audio and semantic embeddings,'' in \emph{2020 IEEE International Symposium on Multimedia (ISM)}.\hskip 1em plus 0.5em minus 0.4em\relax IEEE, 2020, pp. 41--48.

\bibitem{chen2022interactive}
C.~Chen, N.~Hou, Y.~Hu, H.~Zou, X.~Qi, and E.~S. Chng, ``Interactive audio-text representation for automated audio captioning with contrastive learning,'' \emph{arXiv preprint arXiv:2203.15526}, 2022.

\bibitem{koh2022automated}
A.~Koh, X.~Fuzhao, and C.~E. Siong, ``Automated audio captioning using transfer learning and reconstruction latent space similarity regularization,'' in \emph{ICASSP 2022-2022 IEEE International Conference on Acoustics, Speech and Signal Processing (ICASSP)}.\hskip 1em plus 0.5em minus 0.4em\relax IEEE, 2022, pp. 7722--7726.

\bibitem{liu2021cl4ac}
X.~Liu, Q.~Huang, X.~Mei, T.~Ko, H.~L. Tang, M.~D. Plumbley, and W.~Wang, ``Cl4ac: A contrastive loss for audio captioning,'' \emph{Proceedings of the Detection and Classification of Acoustic Scenes and Events 2021 Workshop (DCASE 2021)}, 2021.

\bibitem{xu2020crnn}
X.~Xu, H.~Dinkel, M.~Wu, and K.~Yu, ``A crnn-gru based reinforcement learning approach to audio captioning.'' in \emph{DCASE}, 2020, pp. 225--229.

\bibitem{mei2021encoder}
X.~Mei, Q.~Huang, X.~Liu, G.~Chen, J.~Wu, Y.~Wu, J.~Zhao, S.~Li, T.~Ko, H.~L. Tang \emph{et~al.}, ``An encoder-decoder based audio captioning system with transfer and reinforcement learning for dcase challenge 2021 task 6,'' \emph{DCASE2021 Challenge, Tech. Rep, Tech. Rep}, 2021.

\bibitem{sun2022automated}
J.~Sun, X.~Liu, X.~Mei, M.~D. Plumbley, V.~Kilic, and W.~Wang, ``Automated audio captioning via fusion of low-and high-dimensional features,'' \emph{arXiv preprint arXiv:2210.05037}, 2022.

\bibitem{koizumi2020transformer}
Y.~Koizumi, R.~Masumura, K.~Nishida, M.~Yasuda, and S.~Saito, ``A transformer-based audio captioning model with keyword estimation,'' \emph{arXiv preprint arXiv:2007.00222}, 2020.

\bibitem{xu2017learning}
J.~Xu, T.~Yao, Y.~Zhang, and T.~Mei, ``Learning multimodal attention lstm networks for video captioning,'' in \emph{Proceedings of the 25th ACM international conference on Multimedia}, 2017, pp. 537--545.

\bibitem{iashin2020multi}
V.~Iashin and E.~Rahtu, ``Multi-modal dense video captioning,'' in \emph{Proceedings of the IEEE/CVF conference on computer vision and pattern recognition workshops}, 2020, pp. 958--959.

\bibitem{huang2024mavil}
P.-Y. Huang, V.~Sharma, H.~Xu, C.~Ryali, Y.~Li, S.-W. Li, G.~Ghosh, J.~Malik, C.~Feichtenhofer \emph{et~al.}, ``Mavil: Masked audio-video learners,'' \emph{Advances in Neural Information Processing Systems}, vol.~36, 2024.

\bibitem{kim2019audiocaps}
C.~D. Kim, B.~Kim, H.~Lee, and G.~Kim, ``Audiocaps: Generating captions for audios in the wild,'' in \emph{Proceedings of the 2019 Conference of the North American Chapter of the Association for Computational Linguistics: Human Language Technologies, Volume 1 (Long and Short Papers)}, 2019, pp. 119--132.

\bibitem{dosovitskiy2020image}
A.~Dosovitskiy, L.~Beyer, A.~Kolesnikov, D.~Weissenborn, X.~Zhai, T.~Unterthiner, M.~Dehghani, M.~Minderer, G.~Heigold, S.~Gelly \emph{et~al.}, ``An image is worth 16x16 words: Transformers for image recognition at scale,'' \emph{arXiv preprint arXiv:2010.11929}, 2020.

\bibitem{wu2016google}
Y.~Wu, M.~Schuster, Z.~Chen, Q.~V. Le, M.~Norouzi, W.~Macherey, M.~Krikun, Y.~Cao, Q.~Gao, K.~Macherey \emph{et~al.}, ``Google's neural machine translation system: Bridging the gap between human and machine translation,'' \emph{arXiv preprint arXiv:1609.08144}, 2016.

\bibitem{sun2023dual}
J.~Sun, X.~Liu, X.~Mei, V.~K{\i}l{\i}{\c{c}}, M.~D. Plumbley, and W.~Wang, ``Dual transformer decoder based features fusion network for automated audio captioning,'' \emph{arXiv preprint arXiv:2305.18753}, 2023.

\bibitem{kim2023prefix}
M.~Kim, K.~Sung-Bin, and T.-H. Oh, ``Prefix tuning for automated audio captioning,'' in \emph{ICASSP 2023-2023 IEEE International Conference on Acoustics, Speech and Signal Processing (ICASSP)}.\hskip 1em plus 0.5em minus 0.4em\relax IEEE, 2023, pp. 1--5.

\bibitem{devlin2018bert}
J.~Devlin, M.-W. Chang, K.~Lee, and K.~Toutanova, ``Bert: Pre-training of deep bidirectional transformers for language understanding,'' \emph{arXiv preprint arXiv:1810.04805}, 2018.

\end{thebibliography}

\end{document}